\documentclass[proceedings]{JHEP}

\usepackage{epsfig}

\newcommand{\cq}{\mbox{$c$}}
\newcommand{\bq}{\mbox{$b$}}
\newcommand{\jp}{J^P}
\newcommand{\gevcc}{\mbox{GeV/$c^2$}}
\newcommand{\mevcc}{\mbox{MeV/$c^2$}}
\newcommand{\dsprime}{D^{*\prime}}
\newcommand{\dsprimep}{D^{*\prime+}}
\newcommand{\dsp}{D^{*+}}
\newcommand{\bss}{B_J^{*}}
\newcommand{\dss}{D_J^{*}}
\newcommand{\dssz}{D_J^{*0}}
\newcommand{\LQCD}{\mbox{$\Lambda_{\mathrm{QCD}}$}}

\conference{Heavy Flavours 8, Southampton, UK, 1999}

\title{Spectroscopy of excited \bq\ and \cq\ states}

\author{V. Ciulli \\
  European Laboratory for Particle Physics (CERN), 1211 Geneva 23,
  Switzerland \\ Email: \email{Vitaliano.Ciulli@cern.ch} }

\abstract{Recent results on the spectroscopy of excited \bq\ and \cq\ states
  are presented. In particular, these include the first observation of the 
  $D_1$ (light quark spin $j=1/2$) resonance, searches for radially excited
  $\dsprime$ and observations of orbitally
  excited $\bss$ states. The current experimental
  status on excited charmed baryons is also briefly reviewed.}

\begin{document}

The spectroscopy of hadrons containing a $b$ or a $c$ quark is greatly
simplified by the fact that heavy-quark masses are much greater than
the energy scale \LQCD\ of strong interactions.  
In the limit of infinite heavy-quark mass, heavy-light hadron physics can be
described by an effective theory (HQET), which is  
invariant under changes of the flavour and spin of the
heavy-quark~\cite{isgur89,eichten90,georgi90}. 
The consequences of this heavy-quark symmetry (HQS) for spectroscopy
have been worked out in~\cite{isgur}.
The spin of the heavy quark and the total
angular momentum $j$ of the light degrees of freedom are separately
conserved. Therefore hadronic states can be
classified by the quantum numbers of the light quarks.
Because of the spin symmetry, for fixed $j \not= 0$,
there is a doublet of degenerate states with total spin
$J=j\pm\frac{1}{2}$, and their total 
widths for a strong transition to an other doublet $j^{\prime}$ are
the same. In addition the partial widths of the four possible
transitions are also predicted by the symmetry. Finally, because of the
flavour symmetry, all of the mass splittings and partial decay widths
of these states are independent of the heavy-quark flavour. 

Since the quark masses are not infinite, HQS is an approximate
symmetry and corrections of order $\LQCD/m_Q$ turn out to be
important. HQET provides a framework for treating these
corrections, which break the degeneracy of the levels.
Therefore spectroscopic data are of invaluable importance to test the
predictions of heavy-quark effective theory. 

\section{D mesons spectroscopy}

The expected spectroscopy for $D$ mesons is summarized in
figure~\ref{f:dspectr}.

\EPSFIGURE{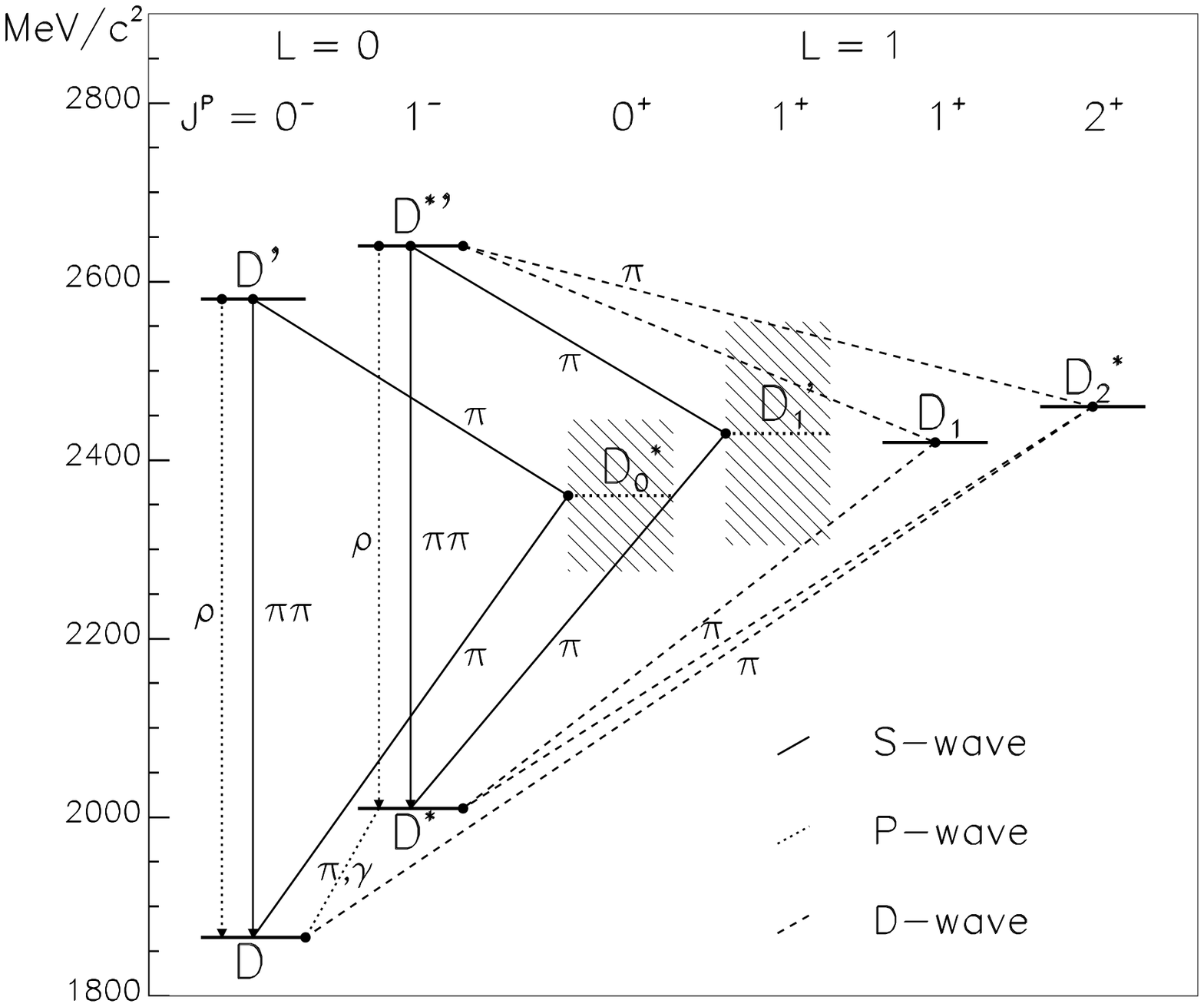,width=7.2cm}{\label{f:dspectr}Spectroscopy
  of D mesons.}

The $D$ and $D^*$ mesons, in which the light quark has
orbital angular momentum $L=0$, are well established and
their properties and decays have been extensively studied~\cite{pdg}.
  
Excited \cq\ mesons with orbital angular momentum $L=1$, collectively
called $\dss$, are classified into two spin
doublets with light-quark angular momentum $j=\frac{1}{2}$ and
$j=\frac{3}{2}$.   
According to HQS, the latter can only decay to $L=0$ states through a
D-wave and they are therefore expected to be narrow. The
$j=\frac{1}{2}$ are instead expected to be wide 
resonances since they can decay via a S-wave. The observed $D_1(2420)$
and $D_2^*(2460)$ are interpreted as the spin 
doublet $j=\frac{3}{2}$, with $\jp=1^+$ and $\jp=2^+$ respectively~\cite{pdg}. 
The first observation of the wide $D_1^{*0}$ has been recently reported by
CLEO II.
It must be noted that a recent calculation~\cite{ebert,isgur98},
which takes into account relativistic effects for the light quark,
predict the mass difference between $D_2^*$ and
$D_1^*$ to be negative ({\em spin-orbit inversion}).
Therefore the observation of the $j=\frac{1}{2}$ doublet is important
not only to confirm general HQS predictions, but also to give insight
into this kind of effect.

A still controversial issue is the observation of the radial excitation
$\dsprime$ claimed by DELPHI, for similar analyses from OPAL and CLEO
found no evidence. 

In the strange sector of charmed mesons, in addition to the $L=0$
states, $D_s^{\pm}$ and $D_s^{*\pm}$, two narrow resonances, the
$D_{s1}^{\pm}(2536)$ and the $D_{sJ}^{\pm}(2573)$, compatible with the
$L=1\, (j=\frac{3}{2})$ doublet have been found~\cite{pdg}, but no new data are
available. 

\subsection{Observation of $D_1^{*0}$ by the CLEO experiment}
The first observation of the $D_1^{*0}$($j=\frac{1}{2}$) meson has
been recently reported by CLEO II~\cite{cleo, dscleo}. Using $3.1\ fb^{-1}$
taken at the $\Upsilon$(4S), they searched for the $\dss$ in the decay
$B^-\to \dssz\pi^-$, followed by $\dssz\to D^{*+}\pi^-$ and $D^{*+}
\to D^0\pi^+$. By measuring the momenta of the three pions and using
four-momentum conservation, the momenta of all particles can be constrained
without any attempt to reconstruct the $D^0$ decay.
This greatly enhances the statistical resolution, although it also
increases the overall level of background. 

\TABULAR{lccc}{
  \hline
  & $m_{D_1^{*0}}$(MeV) & $m_{D_1^0}$(MeV) &
  $\Delta m$(MeV) \\ \hline
  measured   & $2461\pm53$  & $2422.2\pm1.8$\cite{pdg} & $+39^{+42}_{-35}$ \\
  \cite{godfrey} &   $2460$ & $2470$ &$-10$  \\
  \cite{ebert}  &   $2501$ & $2414$ &$+87$  \\
  \cite{isgur98} &   $2585$ & $2415$ &$+170$  \\\hline
  }{\label{t:broad} Comparison of CLEO preliminary result on
  $D_1^{*0}$ mass with theoretical predictions.}

\EPSFIGURE{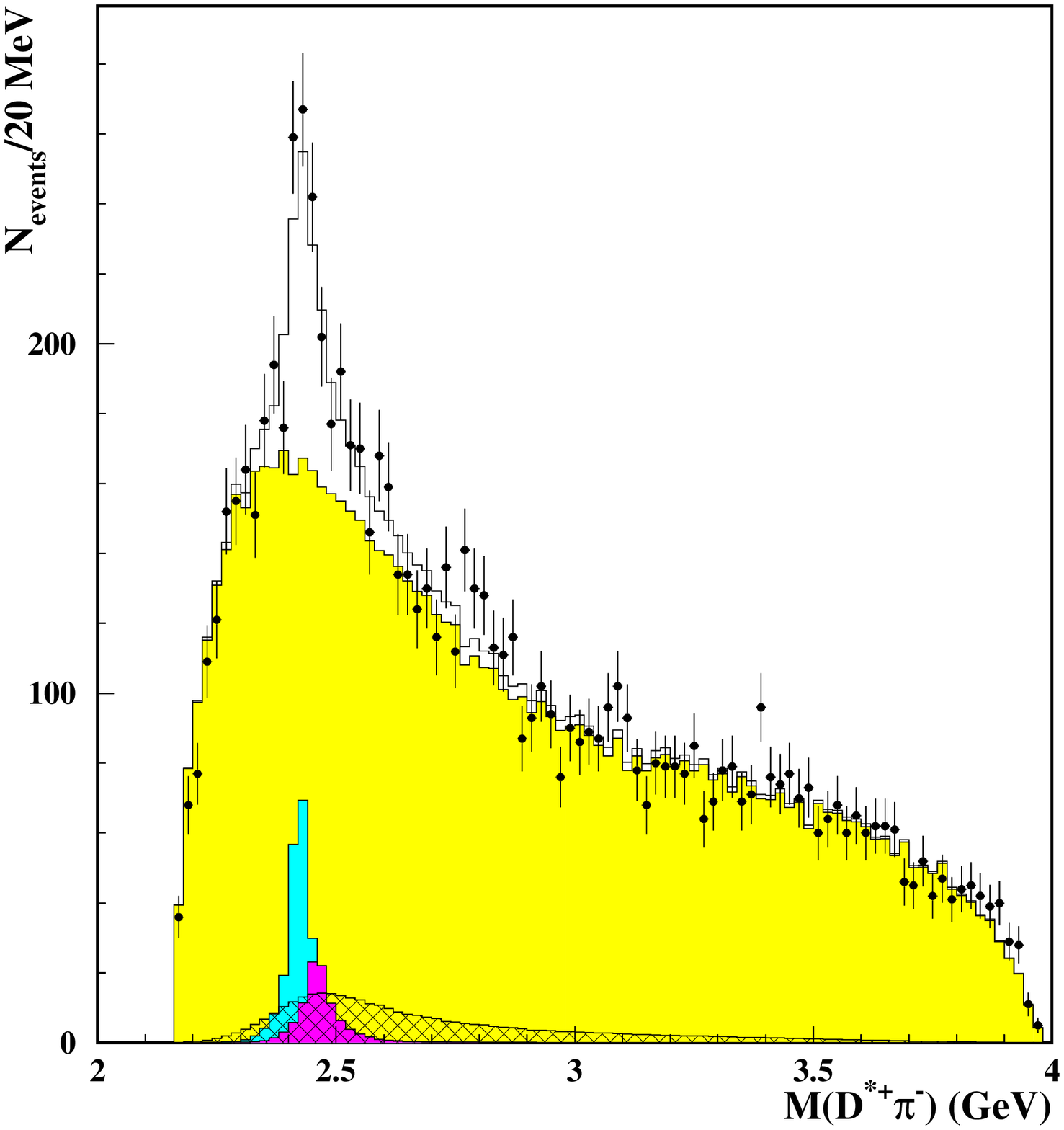,width=7.2cm}{\label{f:cleo} Mass
  spectrum of $\dss$ candidates. The fitted resonances (two narrow and
  one wide) are shown.}

As shown in figure~\ref{f:dspectr}, three of the $\dss$ can decay to
$D^*\pi$. The $D_2^*$ and the $D_1$ through D-wave, and the $D_1^*$
through S-wave. Since both the $B$ and the $\pi$
are pseudoscalar, the $\dss$ is totally polarised and the
relative contributions can be determined from a partial wave analysis of
its decay. A four-dimension fit is performed to the helicities and the
relative azimuth in $\dssz\to D^{*+}\pi^-$ and $D^{*+}\to D^0\pi^+$ decays, and
the $D^{*}\pi$ invariant mass (shown in figure~\ref{f:cleo}). In the
fit, the two $\jp=1^+$ states are allowed to mix, having angular
contributions for both S-wave and D-wave decay, with a strong
interaction phase difference in this interference term. The fit yields  
\begin{eqnarray*}
m_{D^{*0}_1} &=& (2461^{+41}_{-34}\pm10\pm32)\; \mathrm{MeV}\ , \\
\Gamma_{D^{*0}_1} &=& (290^{+101}_{-79}\pm26\pm36)\; \mathrm{MeV}\ .
\end{eqnarray*}
where the first and second errors are respectively statistical and systematic,
while the third one is due to the different chosen 
parametrisation of the strong phases. 
The fit favours $\jp=1^+$ by 2$\sigma$ with
respect to the closest alternative ($0^-$).

In table~\ref{t:broad} the measured $D_1^{*0}$ mass is compared with
the theory. The result is in agreement with HQS
theory, but does not yet allow discrimination between different
models (even if \cite{isgur98} seems to be disfavoured). 

\subsection{Searches for the radial excitation $\dsprime$} 

Two charm radial excitations, a pseudoscalar and a vector state,
respectively called $D^\prime$ and 
$\dsprime$, are expected to exist with masses 
$m_{D^\prime}\approx 2.58$ \gevcc\ and $m_{\dsprime}\approx 2.63$ \gevcc. 
The $\dsprime$ is expected to decay in $D^{*} \pi \pi$, while the
$D^\prime$ decays in $D\pi\pi$, with estimated decay widths of the
order of a few $\mevcc$. 

Evidence of the decay $\dsprimep \to D^{*+}\pi^+\pi^-$ has been
reported by DELPHI~\cite{delphi98}. The $\dsp$ is reconstructed in
the decay mode $\dsp \to D^0 \pi^+$, with either $D^0 \to K^-
\pi^+$ or $D^0 \to K^-\pi^+\pi^+\pi^-$. 
A narrow peak is observed in
the $\dsp\pi^+\pi^-$ invariant mass distribution, as shown in
figure~\ref{f:dspdelphi}, whose width is
compatible with the detector resolution. A signal of $66\pm14$ events,
corresponding to $4.7\sigma$ significance, is obtained with an
observed mass of $2637\pm2\pm6$ \mevcc\ and upper 
limit of 15 \mevcc\ at 95\% C.L. on the full decay width. 
The corresponding production rate, relative to the 
orbitally excited $D_1^0$ and $D_2^{*0}$ decaying to $\dsp\pi^-$, is  
$$R = \frac{n(\dsprimep \to \dsp
  \pi^+\pi^-)}{n(D_1^0,D_2^{*0}\to \dsp\pi^-)} = 0.49\pm0.21\ .$$
The production rate is about the same in $c\bar{c}$ and $b\bar{b}$ events.
  
\EPSFIGURE{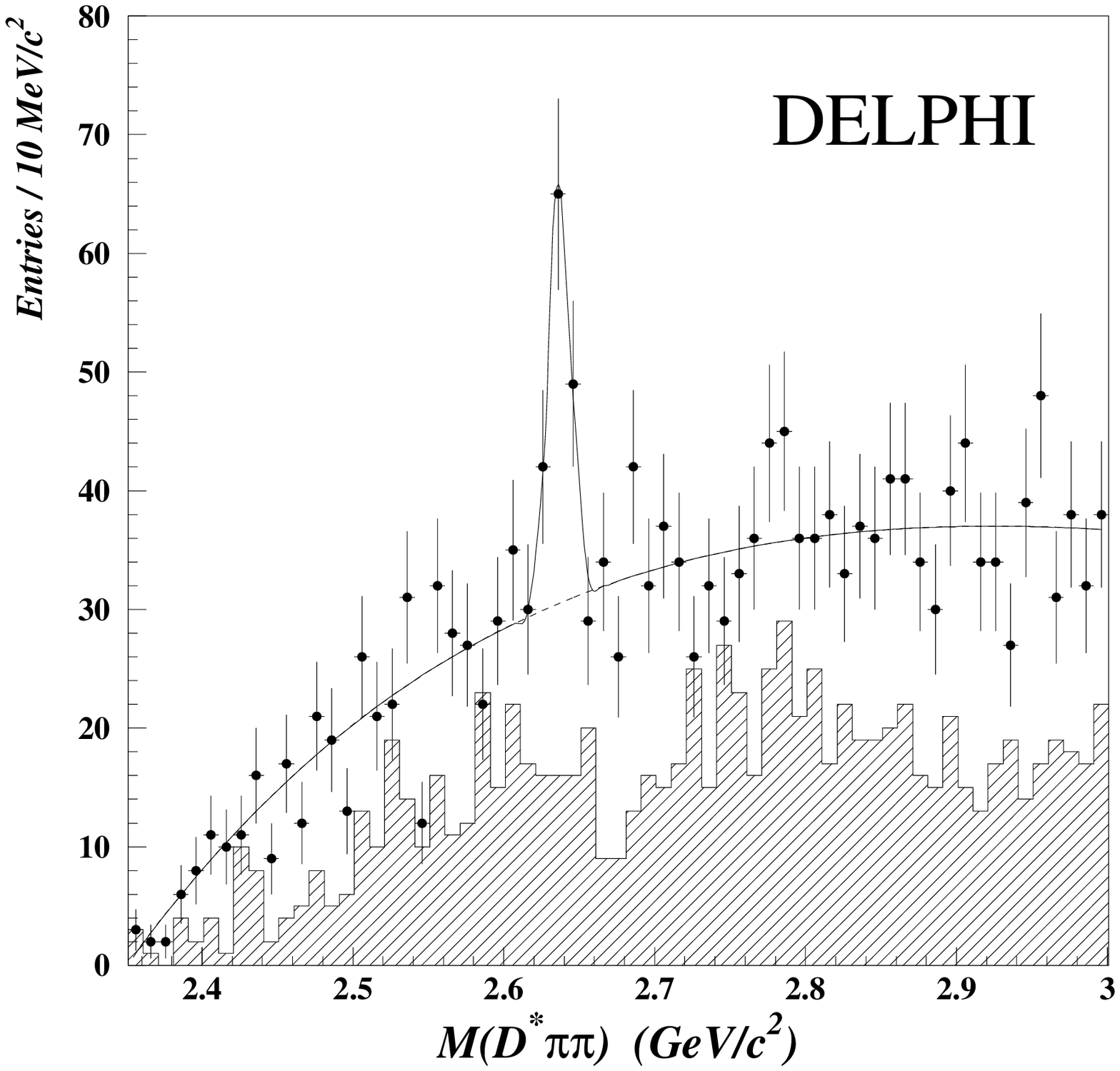,width=7.2cm}{\label{f:dspdelphi}
  The $\dsprime$ signal found by DELPHI (points).
  The hatched histogram shows the
  $D^{*+}\pi^-\pi^-$ combinations.
}
 
Searches for the decay $\dsprimep\to\dsp\pi^+\pi^-$ have been
performed by  
OPAL~\cite{dopal} and CLEO~\cite{dscleo} too.
OPAL use only $D^0\to K^-\pi^+$ but claims a sensitivity similar to
DELPHI.  No evidence for a narrow
resonance is found, yielding a limit $R<0.21$ (95\% C.L.). 
Figure~\ref{f:dspopal} shows the $\dsp\pi^+\pi^-$ invariant 
mass distribution in data and in Monte Carlo, with the
expected signal assuming DELPHI result.
No signal was found by CLEO either, who used $D^0\to
K^-\pi^+\pi^0$ in addition to the decay modes used
by DELPHI. The $\dsp\pi^+\pi^-$ invariant mass is shown in
figure~\ref{f:dspcleo}. Again no evidence for a signal is found and
the limit set on the production relative
to orbitally excited mesons is $R<0.10$ (95\% C.L.). 

\EPSFIGURE{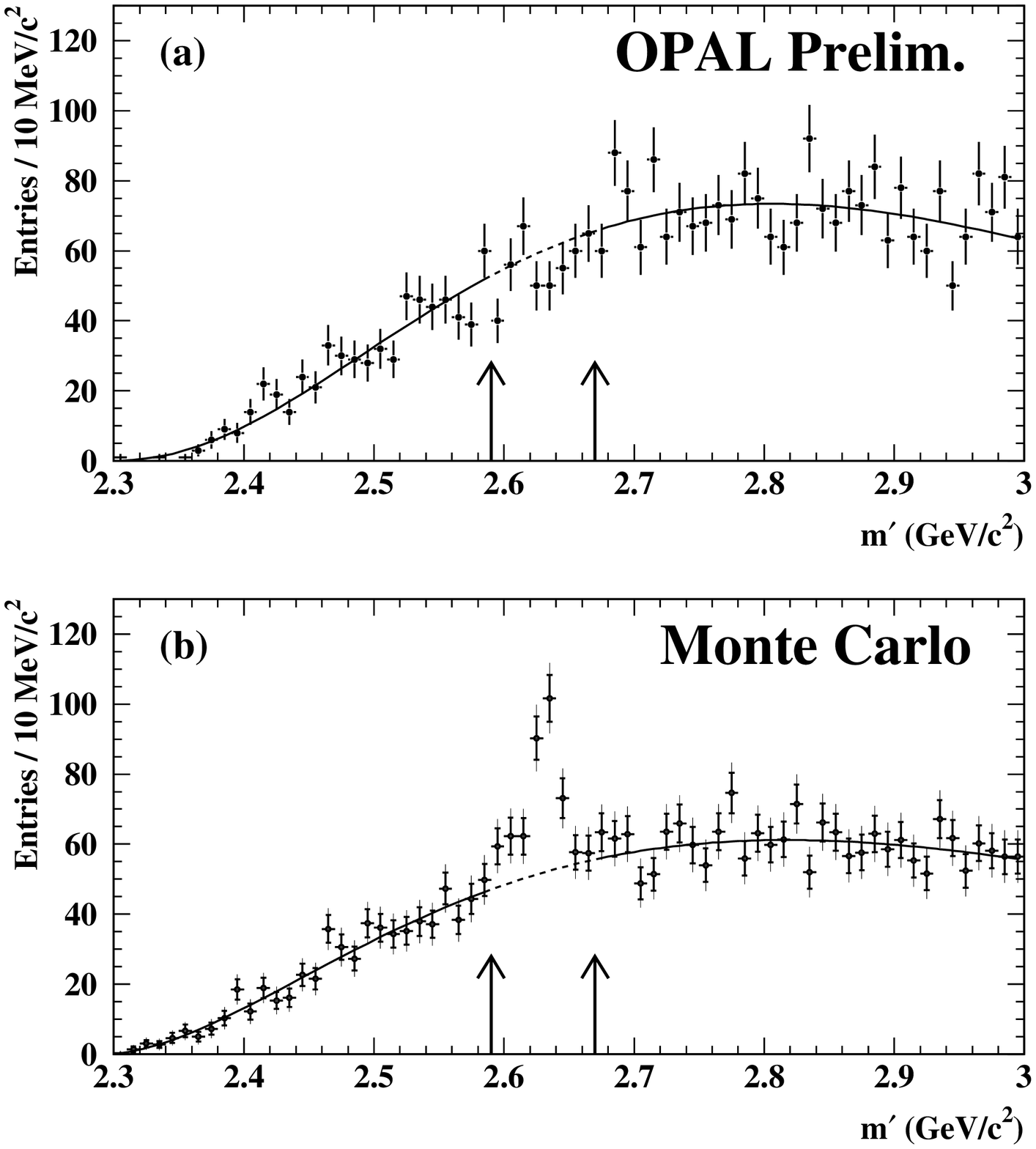,width=7.2cm}{\label{f:dspopal}
  OPAL invariant mass distribution of $D^{*+}\pi^-\pi^+$ combinations in the
  data (top) and in the simulation (bottom). The $\dsprime$ production
  rate in the simulated sample is fixed to the value measured by DELPHI.}

\EPSFIGURE{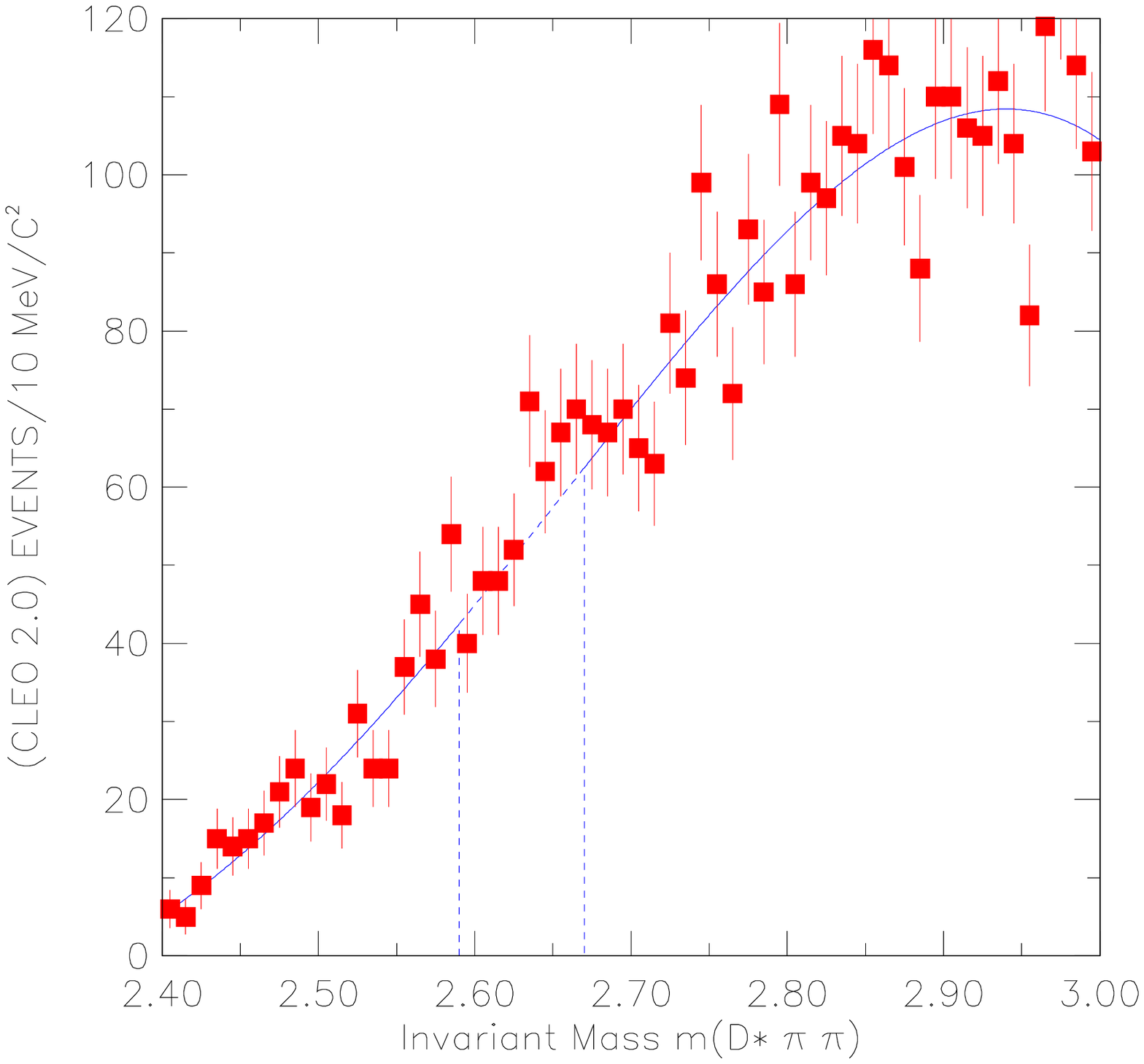,width=7.2cm}{\label{f:dspcleo} Mass of
  CLEO $\dsprimep$ candidates.}

\section{B mesons spectroscopy}

The spectroscopy of B mesons is expected to be very similar to that of
D mesons, but up to now much less information is available compared with
the charm sector.   

The first experimental results on $b$ excited
states came from LEP, where the $B$ mesons are produced in the decays
$Z\to b\bar{b}$. Flavour-charge correlations and a resonant structure
have been observed between the $B$ mesons and nearby pions around an
invariant mass of $5697\pm9$ MeV~\cite{pdg}, being consistent with
orbitally excited $\bss$ mesons decaying to $B^{(*)}\pi$.
Recently the CDF experiment at the Tevatron $p\bar{p}$ collider
also reported a preliminary result on flavour-charge correlations
consistent with the same hypothesis.   

All these analyses are based on inclusive or partial reconstruction of
$B$ mesons, therefore the $B\pi$ mass resolution is only of order of 40 MeV.
This does not allow identification of the $B_2^*$ and $B_1$ narrow resonances
separately, since they are expected to differ only by about 10 MeV. 
In addition, for decays to $B^*\pi$ where the photon in the subsequent decay
$B^*\to B\gamma$ is not detected, the reconstructed $\bss$ mass is shifted
by $-46$ \mevcc.
However several experiments attempted to investigate the
structure and decays of these resonances by using 
different analysis approaches and trying to fit the mass spectrum under
several model assumptions. 

As far as bottom-strange mesons are concerned, no new results are
available on the $B_{sJ}^*(5850)$ resonance~\cite{pdg}, which is
interpreted as coming from orbitally excited $B_s$ mesons. 

\subsection{Observation of orbitally excited $B$ at CDF}

To search for $\bss$, 
CDF~\cite{bcdf} uses $B$ mesons partially reconstructed in semileptonic decays
$B\to\ell\nu D^{(*)}$. This is a natural choice since the lepton
provides an easy trigger for the event. The $D^{(*)}$ meson is fully
reconstructed in several decay modes. Both the production and decay vertices
of the B are reconstructed and used 
to estimate its flight direction, while its momentum is taken from the
decay products, after rescaling by 15\% for the neutrino.
This partial reconstruction also allows the identification
of the charge and flavour
of the $B$ meson.

Once a B is found, it is combined with all pions from the primary
vertex to form $\bss$ candidates. No selection is applied to the pions
in order not to
bias the invariant mass distribution. The analysis
is performed in the variable $Q=m(B\pi)-m(B)-m(\pi)$, where 
the uncertainty on the B mass partially cancels out, and a
50 MeV resolution is achieved.
 
After the selection an excess in $\pi$ charge --- $B$ flavour right-sign
correlations is observed. In order to extract the signal several
background sources have to be considered. These can be divided into
{\em correlated} and {\em uncorrelated} backgrounds, depending on whether
they are produced in association with the $b$ quark (and therefore
carry information on its charge and momentum) or whether they are independent
of the presence of a heavy quark in the event. The latter are 
due to fake B candidates, ``pile-up'' events and particles from the
``underlying'' event and are subtracted from the sidebands in the
selection cuts. The $Q$ distribution after the subtraction of
uncorrelated backgrounds is shown in figure~\ref{f:cdf}. 
\EPSFIGURE{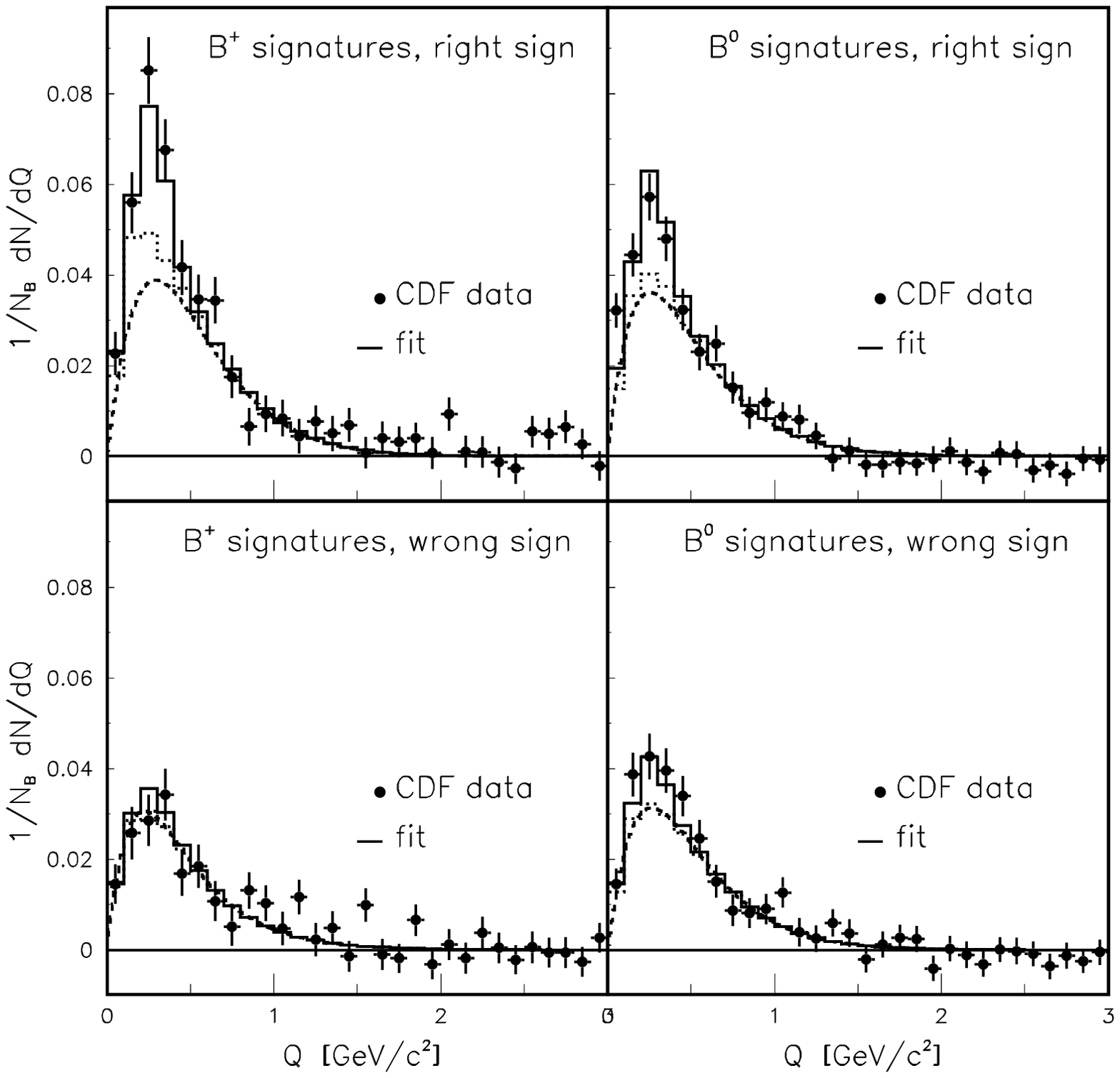,width=7.2cm}{\label{f:cdf} The sideband
  subtracted $B\pi^{\pm}$ $Q$ distributions of the data (points)
  compared to the fit results. The dashed curves are the fitted
  hadronization component, the dotted histograms include all
  backgrounds, and the solid histograms are the totals including
  the $\bss$ signal.}
The remaining correlated background requires a more 
involved treatment, in particular that coming from 
hadronization tracks around the $B$ meson. Pions produced in
association with the $b$ most probably give right-sign
combinations. This is taken into account using a Monte Carlo
inspired parametrization, which is fitted to the data together with 
the signal. Other correlated backgrounds are due to a few
processes involving $B$ mesons, which can be misidentified as $\bss$
decays. The fit to the signal yields
$ {\cal B}(b\to \bss) = 0.28\pm0.06\pm0.03 $, and
assuming the mass splittings between the four $\bss$ as 
in~\cite{isgur98} the mass of the $B_1$($\jp=1^-$,\, $j=\frac{3}{2}$) is 
$$
m_{B_1} = 5.71\pm0.02 \, \gevcc\ .
$$   

\subsection{ALEPH exclusive reconstruction of $\bss$ states}
A different, exclusive approach has been used by ALEPH
in~\cite{aleexcl}, where charged and neutral $B$ mesons are fully
reconstructed and used to study the $B\pi$ resonant structure. This
exclusive approach clearly has the disadvantage of being statistically
limited, but the $B\pi$ mass resolution is improved by one order of
magnitude compared with inclusive analyses, and therefore it may allow
the observation of a narrower structure. Furthermore, the identity and
decay proper time of the $B$ mesons are accurately known and there is
a clear separation between tracks from the $B$ decays and tracks from
fragmentation. Therefore, using this method, one can study the
possibility of tagging the flavour of the $B$ neutral system at production by
the charge of the nearby pion, which has important implications for
the study of CP violation and mixing~\cite{CPviol}. This tagging
capability is clearly enhanced by
the existence of resonant structure in the $B^{(*)}\pi$ system.   

The $B$ mesons are reconstructed in the decay modes $B\to D^{(*)} X$,
where $X$ is a $\pi^{\pm}$, a 
$\rho^{\pm}$ or an $a_1^{\pm}$, and $B\to J/\psi(\psi^{\prime})X$, where
$X$ is a $K^{\pm}$ or a $K^{*0}$. The selection yields 238 $B^+$ and
166 $B^0$, 80\% of the sample being in $B\to D^{(*)} X$ decay mode 
and the remaining
20\% being in $B\to J/\psi(\psi^{\prime})X$. Once a $B$ meson is
reconstructed, nearby pions are classified as ``right-sign'' or
``wrong-sign'' according to their charge. The right-sign track
which has the 
highest momentum along the $B$ direction and, in addition, gives an
invariant mass with the $B$ of less than $7.3$ \gevcc\ is selected to
form a $\bss$ candidate.  
Choosing instead the wrong-sign combination serves as a
check on the estimate of the background and the reliability of the
simulation. The B mass is constrained to its nominal value, yielding
a $B\pi$ invariant mass resolution from 2 to 5 \mevcc\ in the
mass range from $5.5$ to $5.8$ \gevcc.
\EPSFIGURE{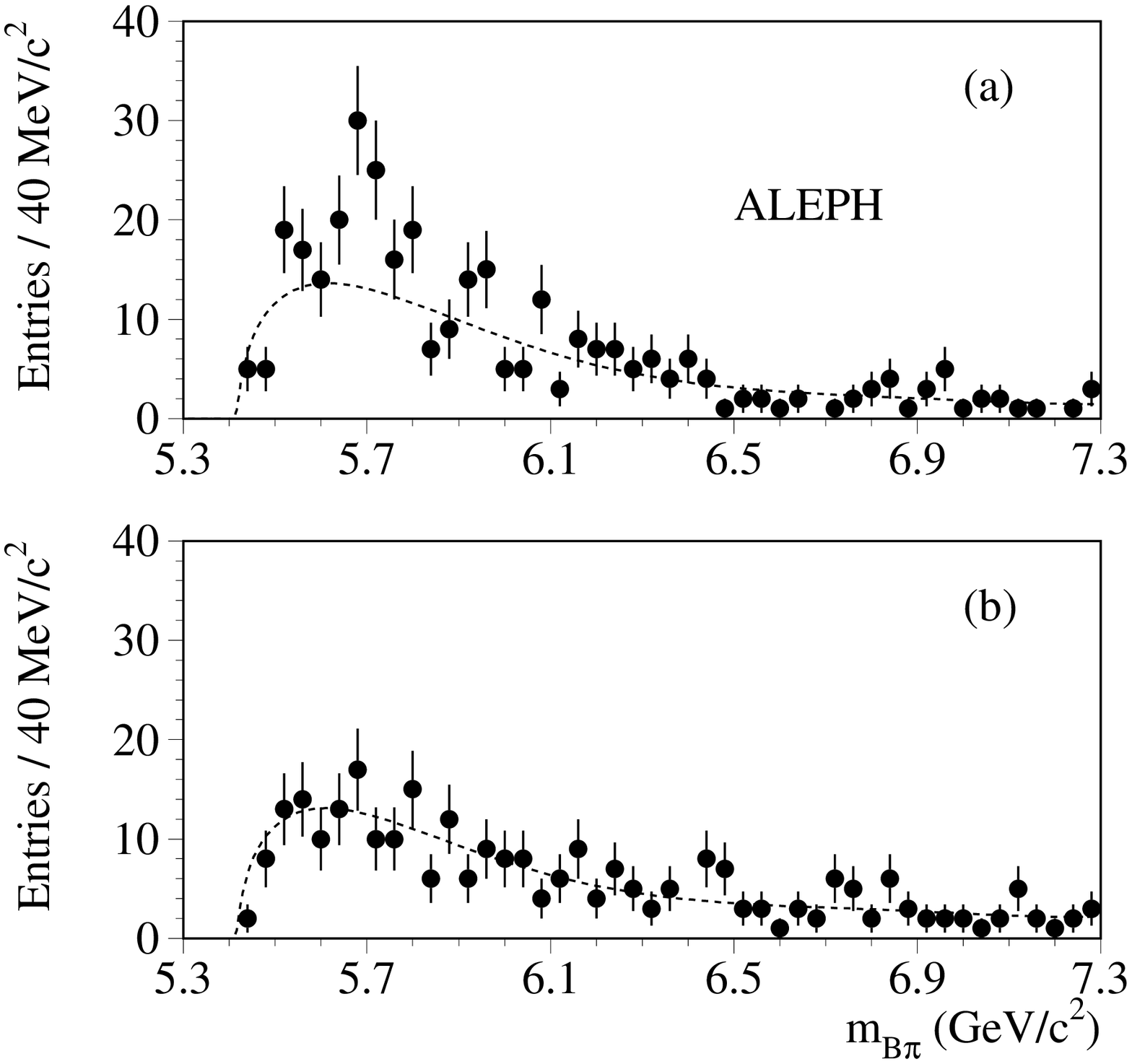,width=7.2cm}{\label{f:alebss} The $B\pi$
  mass distribution in the ALEPH data with the expected background
  shapes for (a) right-sign and (b) wrong-sign candidates.}
The invariant mass  distribution of the selected sample is
shown in figure~\ref{f:alebss} for both right-sign and
wrong-sign combinations. A fit to the signal with a single
Gaussian yields a mass $m_{\bss}=5695^{+17}_{-19}$ \mevcc, a width $\sigma=
53^{+26}_{-19}$ \mevcc\ and a total of $43^{+17}_{-14}$ events. After
correcting for the efficiency to reconstruct the pion and including
the unobserved $B^{(*)}\pi^0$ decays, the production rate is found to be
$$
\frac{{\cal B}(b\to \bss)}{{\cal B}(b\to B_{u,d})} =
(30^{+12}_{-10}\pm3)\%\ . 
$$ 
These results on mass, width and production rate are in agreement with
the results from inclusive analyses.

In the HQS framework, the $\bss$ signal is the sum of the four
different $L=1$ states. The limited statistics do not permit detailed
measurements of all parameters. Nevertheless assuming mass splittings and decay
branching ratios as provided by the theory, the mass of the narrow
doublet (here parametrized by the $B_2^*$ mass) and the overall
production rate of the four states can be fitted to the
data.
\TABULAR{|c|c|c|c|c|c|}{
\hline
      &  Mass   & $\Gamma$ & relative   &                  &         \\  
State &(\mevcc) & (\mevcc) & prod. rate & ${\cal B}(B\pi)$ & ${\cal
  B}(B^*\pi)$  \\\hline
$B_2^*$ & $m_{B_2^*}$        &   $25$   &  $5/12$   & $0.5$ & $0.5$ \\
$B_1  $ & $m_{B_2^*} -12$    &   $21$   &  $3/12$   &   --  & $1.0$ \\
$B_1^*$ & $m_{B_2^*} - 100$  &  $150$   &  $3/12$   &   --  & $1.0$ \\
$B_0^*$ & $m_{B_1^*} -12$    &  $150$   &  $1/12$   & $1.0$ &   --  \\ \hline
}{Properties, relative production rates and branching ratios as used
by ALEPH in the fit to the $B\pi$ invariant mass spectrum.\label{t:aleph}}
Table~\ref{t:aleph} shows the parameters used in the fit.
The $B_2^*$, $B_1$ mass difference and widths, and the equality of the
$B_2^*$ branching ratios to $B^*\pi$ and $B\pi$, are as calculated
in~\cite{eichten93}. The masses and widths of the two wide states
correspond to rough theoretical expectations~\cite{gronau}.
The relative production rates are set according to spin
counting. Finally the
$B\pi$ mass spectra for the $\bss\to B^*\pi$ decays are displaced
downward by 46 \mevcc\ to take into account  the missing soft photon
from the $B^*\to B\gamma$ decay. 
\EPSFIGURE{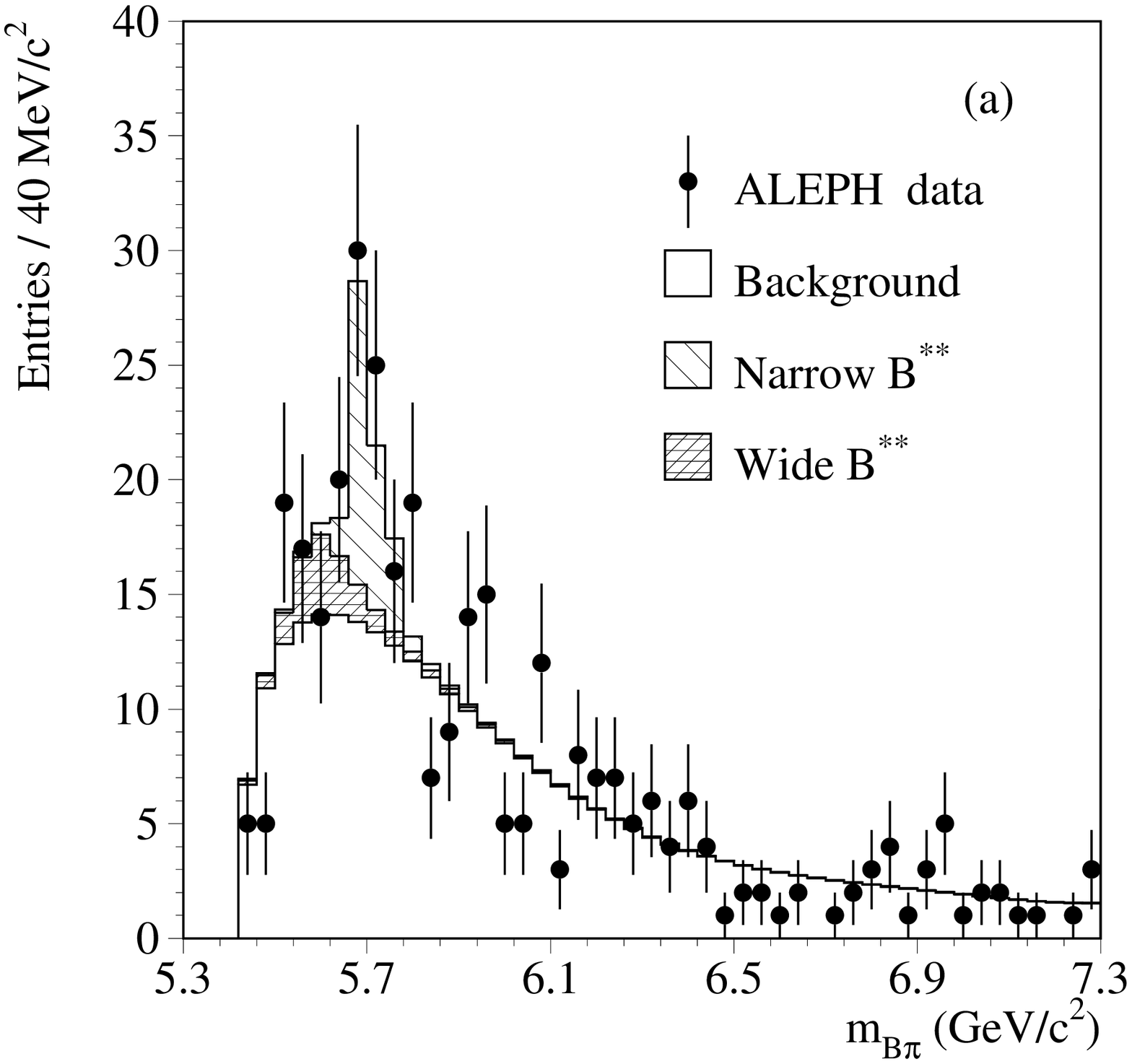,width=7.2cm}{\label{f:alefit} ALEPH 
  right-sign $B\pi$ mass distribution from data with the results of
  the fit.}
Figure~\ref{f:alefit} shows a comparison of the mass
spectrum obtained from the fit and the data. The final values are 
\begin{eqnarray*}
m_{B_2^*} = 5739^{+8}_{-11}(\mathrm{stat})^{+6}_{-4}(\mathrm{syst}) \, \mevcc \\
\displaystyle\frac{{\cal B}(b\to \bss)}{{\cal B}(b\to B_{u,d})} =
(31\pm9(\mathrm{stat})^{+6}_{-5}(\mathrm{syst}))\%\ .
\end{eqnarray*}
The systematic error on the $B_2^*$ mass is dominated by the
uncertainty on the relative production rate, which is estimated by
changing from spin to state counting, while the effect of varying the mass
difference between $B_2^*$ and $B_1^*$ from 50 to 150 \mevcc\ is only
$\pm1$ \mevcc.

\EPSFIGURE{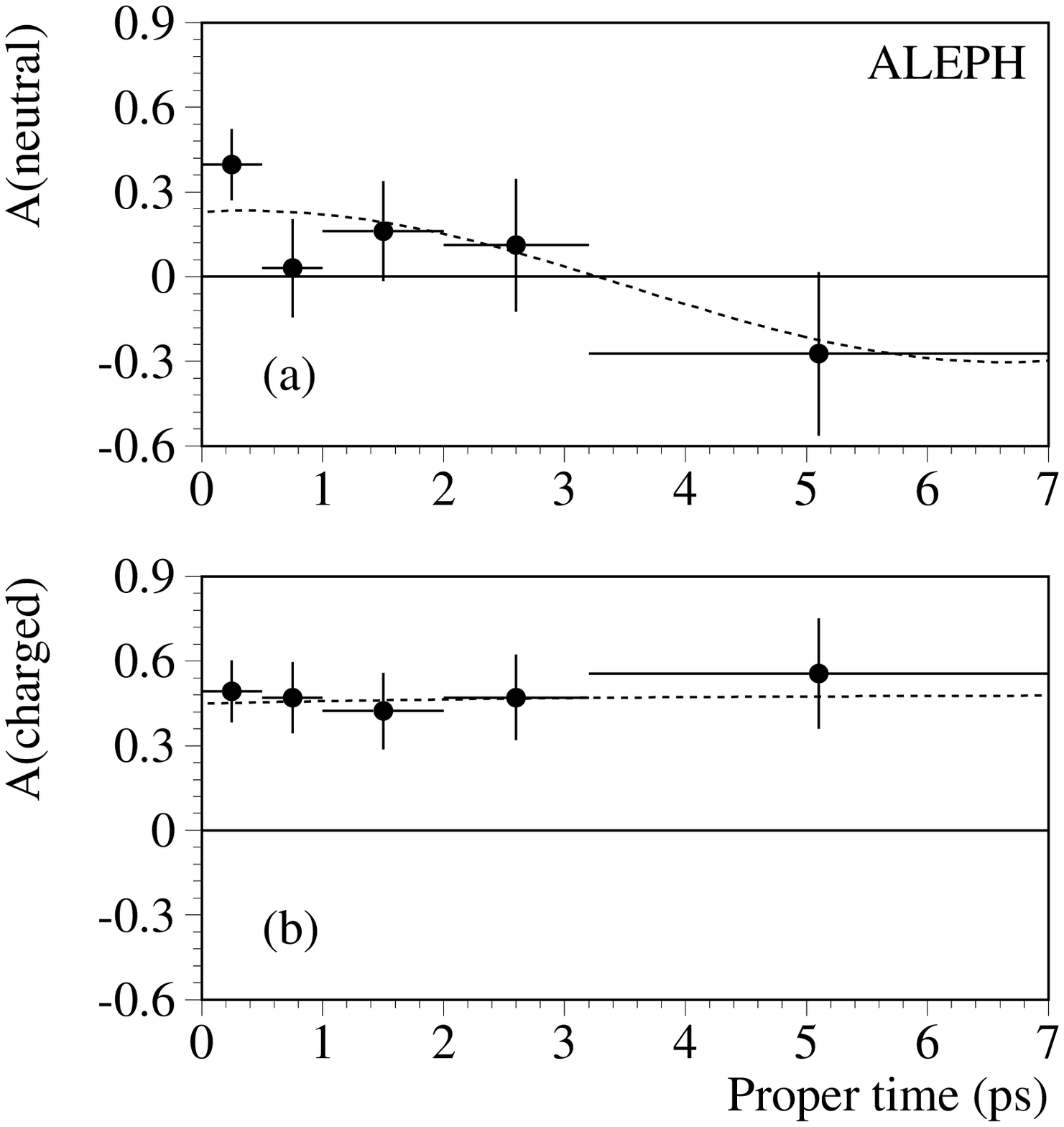,width=7.2cm}{\label{f:aletag} The
right-sign/wrong-sign asymmetry ${\cal A}$ in the data as a function
of the $B$ proper decay time for (a) neutral and (b) charged $B$
mesons. The dashed curves display the results of the fit.}

The same $B\pi$ sample has been used to study the $B$ flavour tagging
based on the pion charge.  
Figure~\ref{f:aletag} shows the asymmetry ${\cal A} = (N_{rs} -
N_{ws})/$ $(N_{rs} + N_{ws})$ as a function of the $B$ decay proper time, for
both neutral and charged $B$ mesons.  
The physical quantity of interest is the mistag rate $\omega^N_{tag}$
for neutral $B$'s. This is extracted from ${\cal A}^N$ taking into
account the effect of $B^0-\bar{B}^0$ mixing. As a check the mistag rate has
also been measured for charged $B$'s, where ${\cal A}^C=(1-2\omega^C_{tag})$,
apart for a small dependence on proper time due to the
different lifetime of the fake $B$'s.  The fit yields
$\omega^N_{tag}=(34.4\pm5.5\pm1.0)\%$ and  
$\omega_{tag}^C=(26.0\pm3.6\pm0.7)\%$. On a high statistics sample of
simulated events the mistag rate for neutral and charged $B$'s are found
to be, respectively, 36\% and 32\%, in good agreement with the above
results. Further studies have shown that the difference in the mistag
rates between neutral and charged $B$ mesons is due to strange quark production
in fragmentation. This spoils the isospin symmetry between the $u$
and $d$ quarks~\cite{dunietz}, since a $K^{\mp}$ is produced in association
with a $B^{\pm}$, carrying flavour information, while a
$\bar{K}^0(K^0)$ results in the fragmentation to    
a $B^0(\bar{B}^0)$. Unless there is perfect pion-kaon
separation the mistag rate is therefore different,
and charged $B$'s cannot be used to infer results for the
neutral $B$'s. 

\subsection{L3 results on $\bss$ spectroscopy}

Preliminary results on the spectroscopy of excited $B$ mesons have
also recently been presented by L3~\cite{bl3}. The mass and width of
$\bss$ have been measured using 1.25 million hadronic Z decays, where
B meson candidates are inclusively reconstructed and combined with
charged pions produced at the primary vertex. 

The B direction is measured using combined information
from the secondary vertex and track rapidity and the B energy is
estimated from a kinematic constrained fit to the centre-of-mass
energy. The obtained resolution on the $B\pi$ mass varies from 20
\mevcc\ to 60 \mevcc, for energies between $5.6$ \gevcc\ and $5.8$ \gevcc.

An excess of events above the expected background, estimated from
the simulation, is found in the region $5.6 - 5.8$ \gevcc\ (see
figure~\ref{f:bl3}). The signal has been fitted assuming a hyperfine
splitting $m_{B_2^*}-m_{B_1}=m_{B_1^*}-m_{B_0^*}=12$ \mevcc, but allowing
both the mass of the $B_2^*$ and of the $B_1^*$ to vary freely. The
widths $\Gamma_{B_2^*}$ and $\Gamma_{B_1^*}$ are also fitted, with the
constraint $\Gamma_{B_1}=\Gamma_{B_2^*}$ and
$\Gamma_{B_0^*}=\Gamma_{B_1^*}$, while the relative production rates are fixed
to spin counting.  The results of the fit are 
\begin{eqnarray*}
m_{B_1^*} &=& (5682\pm23)\, \mevcc \\
\Gamma_{B_1^*} &=& (73\pm44)\, \mevcc \\
 m_{B_2^*} &=& (5771\pm7)\% \\ 
\Gamma_{B_2^*}&=& (41\pm43)\, \mevcc 
\end{eqnarray*}
with a $\chi^2$ of 81 for 73 degrees of freedom. The fit is not a good
description of the data in the region $5.9 - 6.0$ \gevcc\, where an
excess of data events  over the simulated background is present. 
To take into account this excess, a high mass state, which can be
interpreted as a radially excited $B^{\prime}$, has been included in
the fit. 
The final results are  
\begin{eqnarray*}
m_{B_1^*} &=& (5670\pm10\pm13)\, \mevcc \\
\Gamma_{B_1^*}& =& (70\pm21\pm25)\, \mevcc \\
m_{B_2^*} &=&(5768\pm5\pm6)\% \\
\Gamma_{B_2^*} &= &(24\pm19\pm24)\, \mevcc 
\end{eqnarray*}
with a total of $2784\pm274$ events, corresponding to the branching
ratio ${\cal B}(b\to\bss\to B^{(*)}\pi)=0.32\pm0.03$. In addition
$297\pm100$ events occupy the high-mass Gaussian, with 
$m_{B^{\prime}}=5936\pm22$ \mevcc\ and $\sigma_{B^{\prime}}=50\pm23$
\mevcc. Figure~\ref{f:bl3} shows the results of the fit, for which
the $\chi^2$ is 63 for 67 degrees of freedom. 
The differences with respect to the results of the previous fit have
been included in the systematic errors. 
\EPSFIGURE{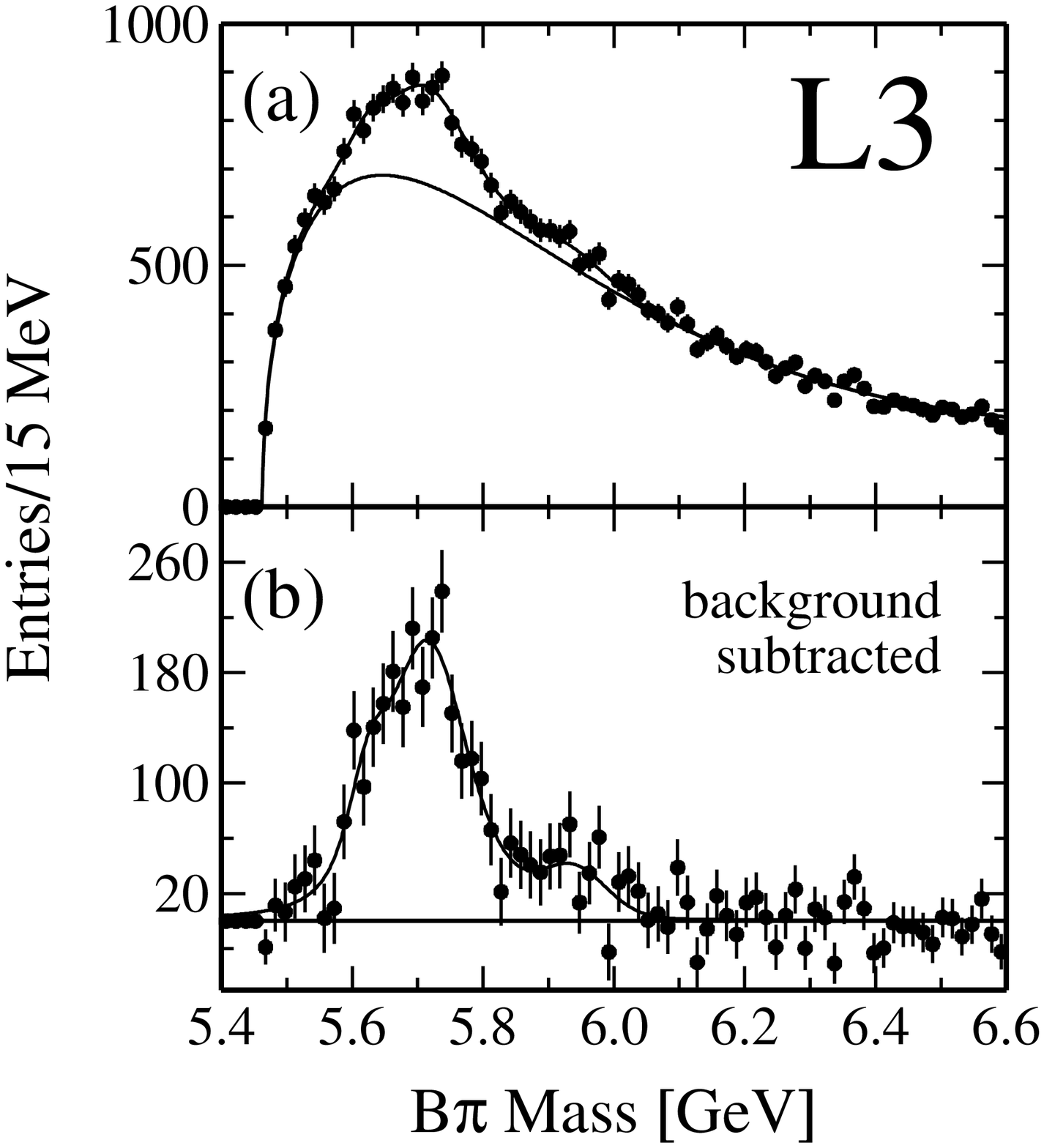,width=7.2cm}{\label{f:bl3} (a) Fit to the
  data $B\pi$ mass distribution. (b) The resulting
  background-subtracted distribution. } 

The above results place the average mass of the $j=3/2$ states
$98\pm11$ \mevcc\ higher than that of the $j=1/2$ states. This supports
an earlier theoretical prediction~\cite{gronau} and disfavours the latest
calculations~\cite{ebert,isgur98}, which predicts spin-orbit
inversion.
To test the ability to discriminate between the two possible cases,
additional fits have been performed, where the mass of the $j=1/2$
states is constrained to be equal to or 100 \mevcc\ higher than the mass
of the $j=3/2$ states. In both cases the fit likelihoods are lower than
before, the highest one being 8.3\% compared to 61.2\% of the
present fit.

\subsection{OPAL study of the $\bss$ decays}   

A preliminary result on ${\cal B} (\bss\to B^*\pi)$ has been recently
presented by OPAL~\cite{bopal}. Using information of the photon in the decay
$B^*\to B\gamma$, OPAL separates the $\bss \to B^*\pi$ decays from the
$\bss \to B\pi$ ones and measures the ${\cal B} (\bss\to
B^*\pi)$. This method gives insight into the decomposition of the
$\bss$ into the states allowed to decay to $B\pi$ ($B_0^*$ and $B_2^*$)
from the other states that can only decay to $B^*\pi$.

\EPSFIGURE{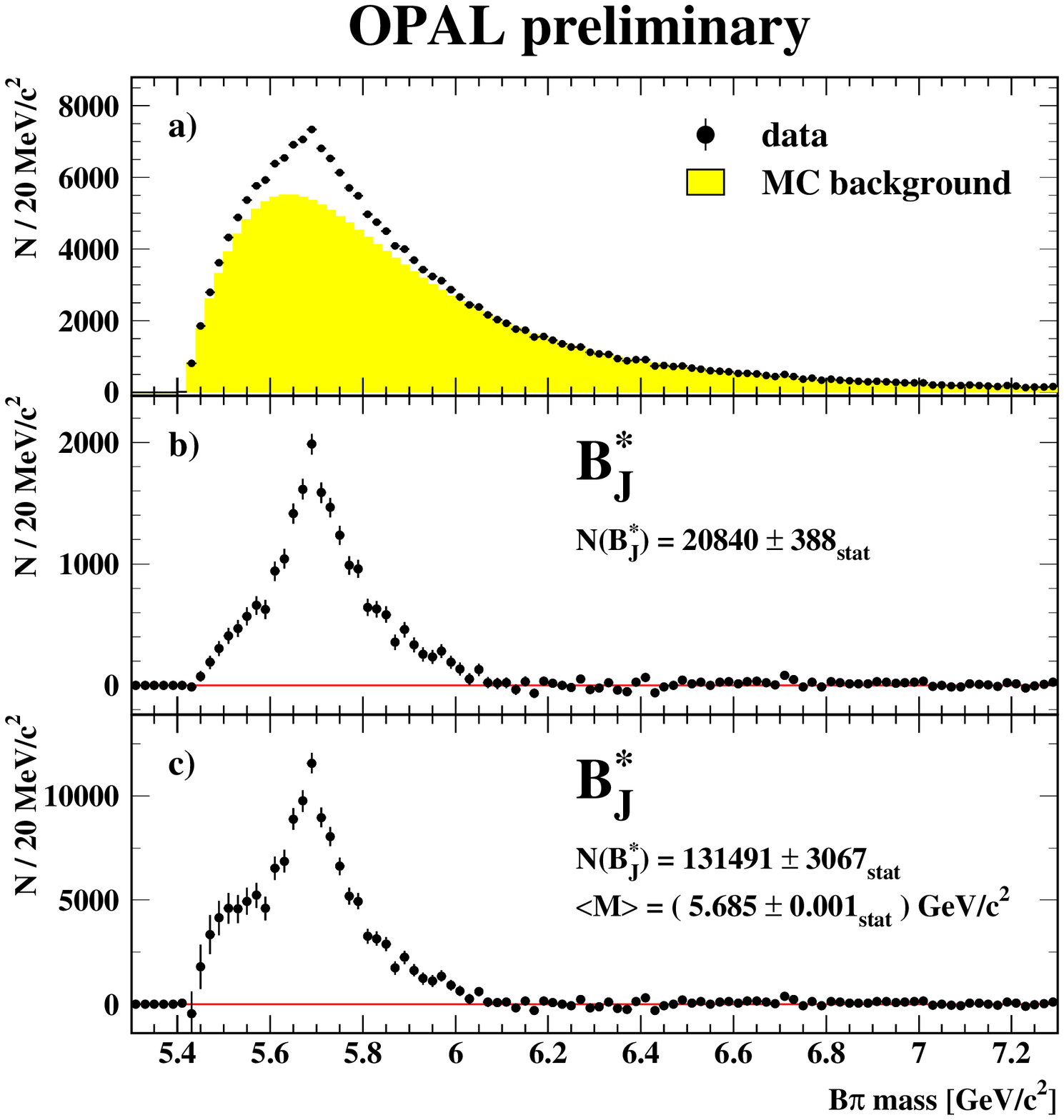,width=7.2cm}{\label{f:opaball} (a)
  The $B\pi$ mass distribution for data. (b) The Monte Carlo
  background-subtracted signal. (c) The efficiency-corrected $\bss$
  signal.} 

The analysis is based on an inclusive reconstruction of $\bss$,
similar to the L3 one previously described. The mass spectrum of the
$B\pi$ sample obtained is shown in figure~\ref{f:opaball}. 
After a candidate $\bss$ is found, photons and converted photons in
the event are used to assess the probability of a $B^*\to B\gamma$
decay. The full sample is then divided into two subsamples, one
enriched and the other depleted in $B^*$, and the ${\cal B} (\bss\to
B^*\pi)$ is extracted from the number of candidates in the two
subsamples. The result is 
$$
{\cal B} (\bss\to B^*\pi) = (85^{+26}_{-27}\pm12)\%\, .
$$
in agreement with theoretical predictions (see table~\ref{t:aleph}).

The composition of the $\bss$ sample is further
investigated by extracting from the two subsamples
the $B\pi$ mass distributions for the two decay modes, which are shown
in figure~\ref{f:opab}. A significant excess of entries is seen in
the pure $\bss\to B^*\pi$ distribution at masses around $5.7$ \gevcc,
with tails down to $5.5$ \gevcc\ and up to $6.0$ \gevcc. The narrow
peak is most likely due to $B_1$ and $B_2^*$ decays, while the tails
may be due to the wide $B_1^*$ state. To
obtain the true mass, the entries have to be shifted to higher mass by
46 \mevcc. In the pure $\bss\to B\pi$ mass distribution, a $2.2\sigma$ excess
is observed at $5.8$ \gevcc, as expected from $B_2^*$ and $B_0^*$ decays. 

\EPSFIGURE{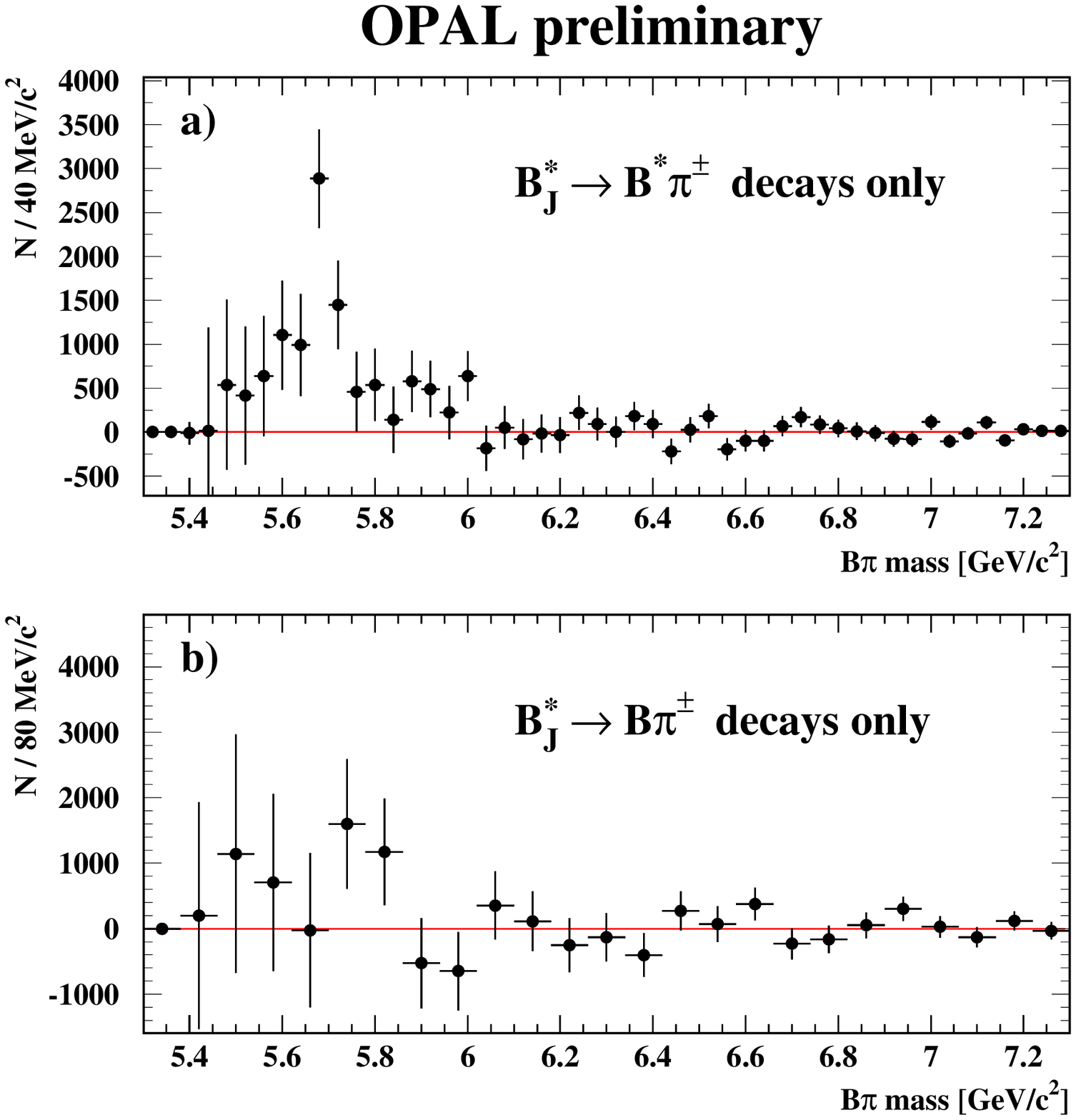,width=7.2cm}{\label{f:opab} The
  $B\pi^{\pm}$ mass distribution of (a) $\bss \to B^*\pi^{\pm}$ and
  (b) $\bss \to B\pi^{\pm}$ transitions.}

\subsection{Summary on $\bss$ states}
 
\TABULAR{llccc}{\hline
& & $m_{B_2^*}$(\mevcc) & $m_{B_1^*}$(\mevcc) & $\Delta m$(\mevcc) \\
\hline
&CDF\cite{bcdf}$^{\ddag}$ & $5725(5745) \pm 20$ & -- & $-160(100)^{\S}$\\
measured &ALEPH\cite{aleexcl} & $5739\pm13$ & -- &$100^{\S}$ \\
&L3\cite{bl3}$^{\ddag}$ & $5768\pm8$ & $5670\pm16$ & $98\pm24$ \\ \hline
&\cite{godfrey}& 5800 & 5780 & $20$ \\
predicted&\cite{ebert} & 5733 & 5757 & $-24$ \\
&\cite{isgur98}& 5715 & 5875 & $-160$ \\\hline
}
{\label{t:bsum} Comparison of experimental results and theoretical
  predictions on $\bss$ masses. The results marked with a $\ddag$ are
  still preliminary. The $\Delta m$, when marked with $\S$, is fixed in
  the fit to the shown value.  }

A summary of present results on the spectroscopy of $b$ excited states is
shown in table~\ref{t:bsum}, together with the predictions of several 
theoretical calculations. ALEPH and CDF results on $B_2^*$ mass are in
agreement between each other and with calculations
in~\cite{ebert}. It must be noticed that they are both rather
insensitive to the assumed mass splitting between the $j=3/2$ and $j=1/2$
doublets: changing the mass splitting from $-160$ \mevcc\ to $100$
\mevcc\ results in an increase of the fitted mass by 20 \mevcc\ in CDF
analysis, while the effect of a similar shift in the ALEPH result is only
few $\mevcc$.
The L3 result on the $B_2^*$ mass is not in good agreement with
ones discussed above, being higher by more than $2\sigma$. This is the only
analysis where the $B_1^*$ mass is also measured and the result, if
confirmed, would strongly disfavour models which predict large spin-orbit
inversion~\cite{isgur98}.  

\section{Charmed baryon spectroscopy}

\EPSFIGURE{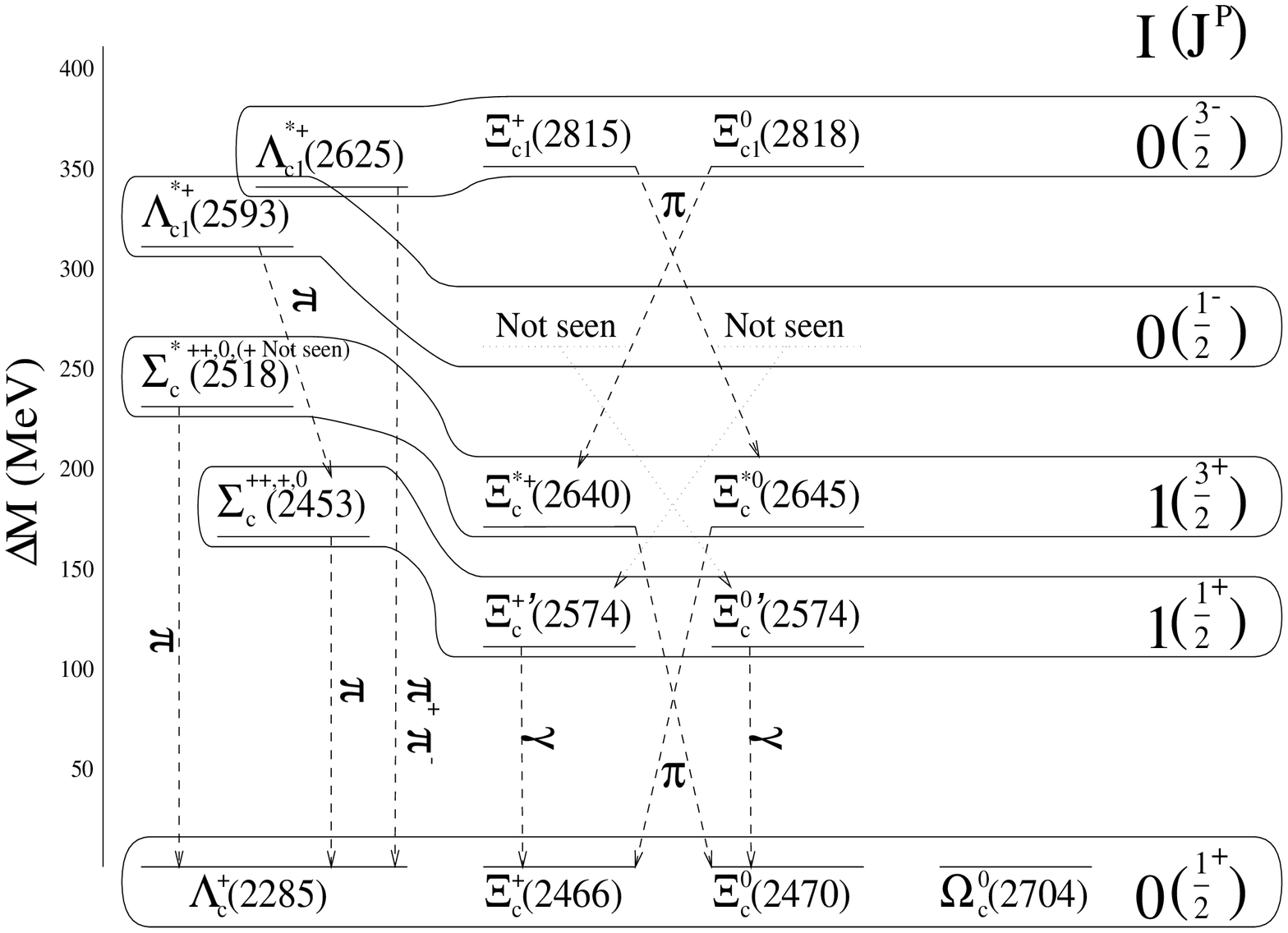,width=11cm}{\label{f:baryon} Spectroscopy
  of charmed baryons}

As far as the charmed baryons are concerned, the most recent results 
concern evidence for the $\Xi_c^{\prime}$ and $\Xi_{c1}$, seen by
CLEO~\cite{xip,xic}.

The $\Xi_c^{\prime}$ are $csq$ baryons (where $q$ can be either a $u$
or a $d$ quark) with the spin of the light quarks $S_{sq}=1$ and
$\jp = \frac{1}{2}^+$. Both charged and neutral $\Xi_c^{\prime}$ are observed
to decay into $\Xi_c\gamma$.

Two new baryons decaying into $\Xi^{+*}_c\pi^-$ and
$\Xi^{*0}_c\pi^-$ have been also observed by CLEO. These new states
have been interpreted as the orbitally excited $\jp=\frac{3}{2}^-$ 
$\Xi_{c1}$ baryons, since the $\jp=\frac{1}{2}^-$ $\Xi_{c1}$,
would decay into $\Xi_c^{\prime}\pi$.

A summary of the present experimental situation on charmed baryons is
shown in figure~\ref{f:baryon}. In the near future new results on
charm baryon spectroscopy are also expected from the fixed target
experiments FOCUS and SELEX. 

\section{Conclusions}

We reviewed recent results on the spectroscopy of excited $b$ and $c$
states. All these results are in agreement with the
expectations from Heavy Quark Symmetry. For the first time 
a wide $j=\frac{1}{2}$ state, namely the $D_1^{*0}$, has been observed.  
However, the observation of the radially excited
$\dsprime$ is still controversial. New studies of $\bss$ resonances also 
support the existence of several states as predicted by HQS.
In particular, if confirmed, both the measured $D_1^{*0}$ mass and the
fits to $\bss$ masses, would disfavour models which
predict large spin-orbit inversion.   


\begin{thebibliography}{10}

\bibitem{isgur89}
N.~Isgur and M.~B. Wise \plb{232}{1989}{113}; \plb{237}{1990}{527}.

\bibitem{eichten90}
E.~Eichten and B.~Hill \plb{234}{1990}{511}.

\bibitem{georgi90}
H.~Georgi \plb{240}{1990}{447}.

\bibitem{isgur}
N.~Isgur and M.~B. Wise \prl{66}{1991}{1130}.

\bibitem{pdg}
C.~Caso {\em et.~al.}, {\it Review of particle physics},  {\em Eur. Phys. J.}
  {\bf C 3} (1998) 1 and 1999 off-year partial update for the 2000 edition
  available on the PDG WWW pages (URL:\href{http://pdg.lbl.gov/}{http://pdg.lbl.gov/}).

\bibitem{ebert}
D.~Ebert, V.~O. Galkin, and R.~N. Faustov \prd{57}{1998}{5663}, [\hepph{9712318}].

\bibitem{isgur98}
N.~Isgur \prd{57}{1998}{4041}.

\bibitem{cleo}
{\bf CLEO} Collaboration, S.~Anderson {\em et.~al.}, {\it Observation of a
  broad \mbox{$L = 1$} \mbox{$c\bar{q}$} state in \mbox{$B^- \to D^{*+}\pi^-
  \pi^-$} at \mbox{CLEO}}, [\hepex{9908009}].

\bibitem{dscleo}
{\bf CLEO} Collaboration, J.~L. Rodriguez, {\it Hadronic decays of beauty and
  charm from \mbox{CLEO}},  [\hepex{9901008}].

\bibitem{godfrey}
S.~Godfrey and R.~Kokoski \prd{43}{1991}{1679}.

\bibitem{delphi98}
{\bf DELPHI} Collaboration, P.~Abreu {\em et.~al.} \plb{426}{1998}{231}.

\bibitem{dopal}
{\bf OPAL} Collaboration, {\it First evidence for a charm radial excitation,
  \mbox{$\dsprime$}},  submitted to {\em XXIX International Conference on
  High Energy Physics, ICHEP 98 (Vancouver), July 23-28, 1998}, 
  OPAL PN 352, July 14, 1998.

\bibitem{bcdf}
{\bf CDF} Collaboration, G.~Bauer, {\it \mbox{CDF B} spectroscopy results: B**
  and \mbox{$B_c^+$}},  [\hepex{9909014}].

\bibitem{aleexcl}
{\bf ALEPH} Collaboration, R.~Barate, {\em et.~al.} \plb{425}{1998}{215}.

\bibitem{CPviol}
J.~Boudreau in {\em Proceedings of the 28th International Conference on High
  Energy Physics, Warsaw, Poland, July 25-31,1996} (Z.~Ajduk and
  A.K.Wr\'oblewski, eds.), p.~1224, World Scientific, 1997.

\bibitem{eichten93}
E.~J. Eichten, C.~T. Hill, and C.~Quigg \prl{71}{1993}{4116},
[\hepph{9308337}]. Updated in FERMILAB-CONF-94-118-T. 

\bibitem{gronau}
M.~Gronau and J.~L. Rosner \prd{49}{1994}{254},
  [\hepph{9308371}].

\bibitem{dunietz}
I.~Dunietz and J.~L. Rosner \prd{51}{1995}{2471},
  [\hepph{9411213}].

\bibitem{bl3}
{\bf L3} Collaboration, {\it Measurement of the spectroscopy of orbitally
  excited \mbox{B} mesons at \mbox{LEP}},  {\em {\rm submitted to}
  International Europhysics Conference High Energy Physics 99, Tampere, July
  15-21,1999, {\rm L3 Note 2423, June 15, 1999}.}

\bibitem{bopal}
{\bf OPAL} Collaboration, {\it Investigation of the decay of orbitally-excited
  \mbox{B} mesons and first measurement of the branching ratio
  \mbox{$BR(\bss\to B^*\pi)$}},  {\em {\rm submitted to} International
  Europhysics Conference High Energy Physics 99, Tampere, July 15-21, 1999,
  {\rm OPAL PN400, July 7, 1999}.}

\bibitem{xip}
{\bf CLEO} Collaboration, C.~P. Jessop {\em et.~al.}
\prl{82}{1999}{492}, [\hepex{9810036}] .

\bibitem{xic}
{\bf CLEO} Collaboration, J.~P. Alexander {\em et.~al.}, {\it Evidence of new
  states decaying into \mbox{$\Xi_c^* \pi$}}, [\hepex{9906013}].

\end{thebibliography}

\end{document}